# Electrical, magnetic, magnetodielectric and magnetoabsorption studies in multiferroic GaFeO$_3$

V. B. Naik and R. Mahendiran[1]

Department of Physics and NUS Nanoscience & Nanotechnology Initiative

(NUSNNI), Faculty of Science, National University of Singapore,

2 Science Drive 3, Singapore -117542, Singapore

**Abstract**

We report electrical, magnetic, magnetodielectric and magnetoabsorption properties of a polycrystalline GaFeO$_3$. The resistivity measurement shows that the sample is highly insulating below 200 K and the resistivity above 200 K obey the Arrhenius law with an activation energy of $E_a$ = 0.67 eV. An anomaly occurs in the temperature dependence of permittivity ($\varepsilon$) near the ferrimagnetic transition temperature ($T_C$ = 228 K) in a zero magnetic field and it is suppressed under $\mu_0 H$ = 60 mT which indicates a possible magnetoelectric coupling in GaFeO$_3$ with a fractional change of $\Delta\varepsilon/\varepsilon$ = -1.8% at 60 mT around $T_C$. The coercivity ($H_C$) of the sample increases dramatically with lowering temperature below 200 K from 0.1 T at 200 K to 0.9 T at 5 K. Magnetoabsorption was studied with a LC resonance technique and we found a close correlation between the shift in the resonance frequency due to applied magnetic field and the coercive field measured using dc magnetization measurements. Our results obtained with multiple techniques suggest that GaFeO$_3$ is an interesting ferrimagnet with potential applications in future multiferroic devices.

---

[1] Corresponding author – phyrm@nus.edu.sg

# I. INTRODUCTION

The magnetoelectric (ME) multiferroic materials which show a strong coupling between ferromagnetic and ferroelectric order parameters are promising for applications in new kind of multistate non volatile memories such as magnetically tunable ferroelectric random access memories (FeRAMs), electrically tunable magnetic random access memories (MRAMs) and high frequency filters etc.[1,2,3,4,5] In addition, the observations of magnetic control of ferroelectric polarization in TbMnO$_3$,[6] electric-field-induced spin flop in BiFeO$_3$,[7] magnetic-field-induced ferroelectric state in DyFeO$_3$[8] and the ME memory effect in MnWO$_4$[9] are quite interesting from both a fundamental and technical perspectives. However, one of the biggest challenges facing the field of multiferroics is the need for room temperature multifunctionality, since there exist a very few single-phase multiferroic materials. Because, ferromagnetism requires an odd number of d-electrons while the ferroelectricity generally occurs only in the absence of d-electrons.[10] This apparent incompatibility to show multiferroicity can indeed be overcome in materials such as BiFeO$_3$, GaFeO$_3$ and rare earth manganites of AMnO$_3$ (A = Y, Tb, Gd, Ho)[11] in which A and Mn are the sources of ferroelectricity and magnetism, respectively.

The magnetic and magnetoelectric properties of GaFeO$_3$ was a topic of intensive research in 1960 since Remeika *et al.*[12] reported the occurrence of piezoelectricity at room temperature above the ferromagnetic transition temperature $T_C$ = 260 K in Ga$_{2-x}$Fe$_x$O$_3$ (x =1). Soon after, Rado *et al.*[13] observed the magnetoelectric effect in Ga$_{2-x}$Fe$_x$O$_3$ ($x \approx 1$) and showed this effect to be larger by an order of magnitude than previously known magnetoelectric material, $Cr_2O_3$.[14] Petrov et al.[15] found that the line shape and the line width of electron spin resonance spectra of the piezoelectric ferrimagnet $Ga_{0.85}Fe_{1.15}O_3$ were affected by an external electrical field and that could not be attributed to change in the conductivity. Folen and Rado suggested[16] the electric field induced change in the electron spin resonance line width observed by Petrov et al.[15] could have been caused by Joule heating. Kaneko et al.[17] found that the $T_C$ in single crystals of $Ga_{2-x}Fe_xO_3$ can be continuously increased from 120 K for x = 0.08 to 370 K for x = 0.14.

$GaFeO_3$ crystallizes in an orthorhombic structure with space group $Pc2_1n$.[18] This compound has a spontaneous polarization along the *b* axis and has a collinear ferrimagnetic structure[19] where the ferrimagnetism is due to the unequal distribution of Fe spins of nearly equal magnitude on the sublattices with a magnetic moment of the spin along the *c* axis (instead of a canted antiferromagnetic structure as previously assumed[13] or inferred[20]). Recent reports of magnetization-induced second harmonic generation,[21] optical[22] and $dc$[23] ME effect, and the ultrafast electric and magnetic response of $GaFeO_3$ induced by irradiation of a femtosecond laser pulse[24] makes this compound more attractive for potential applications. Moreover, $GaFeO_3$ was reported to show the largest Faraday rotation.[25] However, detailed magnetic and electrical studies in $GaFeO_3$ polycrystal have not yet been reported. In this work, we report electrical, magnetic, magnetodielectric and magnetoabsorption properties of a polycrystalline $GaFeO_3$ sample.

## II. EXPERIMENT DETAILS

Polycrystalline sample of GaFeO$_3$ was synthesized using a conventional solid state route. The solid solution was prepared by mixing the stoichiometric mixtures of Ga$_2$O$_3$ and Fe$_2$O$_3$ precursors, and was homogenized in an agate mortar before it reacted at 1000 °C for 12 hours. Final sintering was done at 1200 °C for 12 hours. Single phase identification was performed by the powder X-ray diffraction (XRD) experiment (Philips X'pert Pro) using Cu Kα radiation. The compound was pressed into pellet and heat treated at 1250 °C for 5 hours to get a relatively dense pellet. A parallel plate capacitor structure was made for dielectric and ferroelectric measurements by using the silver paint to make electrodes. The typical dimensions ($A$ – area and $t$ – thickness) of the samples for dielectric and ferroelectric measurements were $A$ = 10.56 mm$^2$, $t$ = 0.91 mm and $A$ = 9.11 mm$^2$, $t$ = 0.38 mm, respectively. Temperature dependence of dielectric properties was studied using the Agilent 4285A Precision LCR meter and an optical cryostat (Janis model CCS102). Two probe resistivity measurement was carried out using an electrometer (Keithley 6517A) with Janis cryostat. A precision LC ferroelectric tester (Radiant Technologies) was used to measure the ferroelectric properties (*P-E* loops). Temperature and field dependences of magnetization measurements were performed using a vibrating sample magnetometer (VSM) with a superconducting magnet and a commercial cryostat (PPMS, Quantum Design Inc). Temperature and magnetic field dependent radio-frequency (*rf*) electromagnetic absorption has been investigated using a home built *LC* resonant circuit powered by an integrated chip oscillator (ICO)[26] with PPMS.

### III. RESULTS AND DISCUSSION

Fig. 1(a) shows the XRD pattern of the GaFeO$_3$ compound at room temperature, which reveals the single phase pattern quite similar to the JCPDS-ICDD 76-1005.[27] The observed XRD pattern (blue color) was indexed (red color) to orthorhombic structure of GaFeO$_3$ with space group *Pc2$_1$n* (*Pna2$_1$*, recommended notation used in the international table) using TOPAS software version 2.1. The prominent peaks in the XRD pattern are indexed by its crystallographic planes, which are represented by the miller indices (*hkl*). The lattice parameters of the compound determined by the above Reitveld refinement are found to be $a$ = 8.74042 Å, $b$ = 9.38931 Å and $c$ = 5.07955 Å which are in close agreement with earlier report.[27] There is no report on the resistivity measurement for GaFeO$_3$ compound, perhaps because the compound is highly insulating. Temperature dependence of the resistivity, $\rho(T)$ is shown in the fig. 1(b) in the temperature interval of 360 K to 265 K. The $\rho(T)$ increases from 50 kΩm at 360 K to 60 MΩm at 265 K. The inset in the fig. 1(b) shows the Arrhenius plot for the resistivity which reveals the linear relationship between ln$\rho(T)$ and 1/*T*. The activation energy deduced from an Arrhenius law, $\rho = \rho_0 \exp(E_a/k_B T)$ is found to be $E_a$ = 0.675 eV.

The fig. 2(a) shows the temperature dependences of the magnetization (*M*) under different magnetic fields, $\mu_0 H$ = 0.2 T, 0.5 T, 1 T, 2 T and 5 T. The rapid increase of magnetization in a magnetic field of 0.2 T at $T_C \approx$ 228 K (calculated from the maximum change in the slope) signals the onset of ferrimagnetic transition in GaFeO$_3$. The magnetization gradually increases with lowering temperature below $T_C$ and becomes nearly temperature independent as it approaches to the lowest temperature $T$ = 10 K. The ferrimagnetic transition broadens with increasing strength of the external magnetic field.

The magnitude of $M$ at 10 K is 0.667 $\mu_B$/Fe atom under the applied field of $\mu_0H$ = 5 T. The sample shows well defined hysteresis loops below 230 K [fig. 2(b)]. The spontaneous magnetization ($M_S$) obtained from the extrapolation of the high field $M$ to $H = 0$ decreases appreciably with increasing temperature. The value of $M_S$ at 5 K is 0.45 $\mu_B$/Fe atom. The significant point here is that the coercive field ($H_C$) is tremendously large which elucidates the hard magnetic material nature of GaFeO$_3$, and it increases rapidly below 200 K from 0.1 T at 200 K to 0.9 T at 5 K [fig. 2(c)].

Temperature dependences of the dielectric constant [$\varepsilon(T)$] and dissipation factor ($D$ = tan$\delta$) in zero field for three different frequencies (100 kHz, 500 kHz and 1 MHz) are shown in the main panels of the fig. 3. The $\varepsilon(T)$ showed an anomaly near ferrimagnetic transition temperature, $T_C \approx 228$ K, but the $D$ does not show any clear anomaly close to $T_C$. Interestingly, a small magnetic field of $\mu_0H$ = 60 mT suppresses the dielectric anomaly found near $T_C$, which is shown in the inset of the fig. 3(a) for $f$ = 1 MHz (left scale). The magnetodielectric (MD) coefficient calculated from the equation MD = [$\varepsilon(H) - \varepsilon(0)$] / $\varepsilon(0)$ is plotted as a function of temperature in the inset of the fig. 3(a) (right scale) and it is found to be -1.8% close to $T_C$ in GaFeO$_3$. The temperature dependence of dissipation factor in $\mu_0H$ = 0 and 60 mT for $f$ = 1 MHz shown in the inset of the fig. 3(b) does not show up any change in response to the applied field as like the $\varepsilon(T)$. In a ferroelectromagnet, the difference in the dielectric constant [$\Delta\varepsilon = \varepsilon(H) - \varepsilon(0)$] below $T_C$ is proportional to the square of the magnetization i.e., $\Delta\varepsilon \sim \gamma M^2$, where $\gamma$ is the magnetoelectric coupling constant.[28] Z H Sun et al.[29] has observed a linear relationship between $\Delta\varepsilon$ and $M^2$ as an indication of the magnetoelectric (ME) coupling, and reported the magnetocapacitance (MC) of -0.5 % close to $T_C$ in GaFeO$_3$ based on the extrapolation of the zero field anomaly. The appearance of dielectric anomaly around $T_C$ in zero magnetic field and its suppression in a small magnetic field of $\mu_0 H$ = 60 mT suggest an active magnetoelectric coupling in our GaFeO$_3$ sample. A similar weak anomaly has been found in $\varepsilon(T)$ in the hexagonal YMnO$_3$[30] at $T_N$ = 70 K. Very recently, T. Kimura *et al.*[31] has found a pronounced anomaly in $\varepsilon(T)$ at the onset of paramagnetic to spiral antiferromagnetic transition ($T_C$ = 270 K) in CuO. It is also to be noted that the observed -1.8% MD coupling coefficient in our sample is higher than -0.6% MC observed in the ferromagnetic BiMnO$_3$[28] at $T_C \approx$ 100 K and $\mu_0 H$ = 9 T, and lower than -8% MD effect found in the *E* type antiferromagnetic HoMnO$_3$[32] at *T* = 4.5 K and $\mu_0 H$ = 7 T.

The *P-E* loops measured at *f* = 1 kHz (frequency of the hysteresis cycle or measurement frequency) at selected temperatures, *T* = 150 K, 200 K and 225 K are shown in the fig. 4(a). The hysteresis loop observed at *T* = 225 K is unsaturated and rounded at the highest field, which reveals the leakage current contribution that overshadows the true polarization due to the orientation of the electric dipoles. Because, for an insulating ferroelectric material, the switched charge *Q* due to applied electric field *E* depends only on the remanent polarization $P_r$ through the relation $Q = 2P_r A$ where *A* is the surface area of the capacitor, whereas for a lossy dielectric material, extra contribution comes from the conductivity $\sigma$ through the relation $Q = 2P_r A + \sigma E A t$, where *t* is the time for hysteresis measurement.[33] The *P-E* loops at *T* = 200 K and 150 K shown in fig. 4(a) are unsaturated even at the higher limit of electric field of 5 kV/cm in our experiment, but it elucidates the considerable suppression of leaky behavior of the compound below $T = 225$ K as supported by our resistivity measurement, where the resistivity increases exponentially as temperature decreases, or in other words, the conductivity ($\sigma$) contribution to the polarization is less at lower temperatures. The highly frequency dependent *P-E* loops at $T = 150$ K [fig. 4(b)] measured in a narrow range of measurement frequency ($f = 1$ kHz – 50 Hz) suggest that our polycrystalline $GaFeO_3$ sample does not appear to be ferroelectric because the *P-E* loops opens up as the measurement frequency decreases which is an artifact due to the leakage current contribution.[33] However, electrical polarization study at higher electric field is necessary to confirm the ferroelectricity in our bulk sample. Next we will see the behavior of magnetic absorption studied by the *LC* resonance technique.

In the main panel of fig. 5, we show the temperature dependences of the resonance frequency ($f_r$) (right scale) and the current (*I*) through an ICO (left scale) for a few *dc* magnetic fields, $\mu_0H = 0$ T, 60 mT and 0.2 T applied along the coil axis. Temperature dependence of $f_r$ and *I* for the empty coil is also shown. The $f_r$ of an ICO in a zero magnetic field ($\mu_0H = 0$ T) decreases gradually from $T = 300$ K, shows a minimum close to the ferrimagnetic transition ($T_C = 228$ K), and then increases gradually with lowering temperature. A small magnetic field of $\mu_0H = 60$ mT appreciably suppresses the minimum in $f_r$ and broadens it, but it hardly affects the $f_r$ in the temperature range $T < 190$ K and T > 250 K. The peak in $f_r$ becomes more broader in a magnetic field of $\mu_0H = 0.2$ T and the behavior of $f_r$ is similar to an empty coil. On the other hand, the current *I* in a zero magnetic field showed a gradual decrease with lowering temperature as like an empty coil, except a weak anomaly showed up very close to $T_C$ in the temperature interval of 200 K to 250 K which can be clearly seen in the inset. This is in contrast to the large decrease in $I$ found at $T_C$ for $La_{0.67}Ba_{0.33}MnO_3$.[26] The weak anomaly present in the zero field $I$ is completely suppressed in the magnetic fields of $\mu_0H$ = 60 mT and 0.2 T.

The fig. 6(a) shows the magnetic field dependence of $f_r$ at selected temperatures $T$ = 10 K – 200 K. These data were taken by monitoring the $f_r$ while sweeping the magnetic field from $\mu_0H$ = 0 T → +3 T → -3 T → +3 T. We have not shown the initial field sweep (0 T → +3 T) here, since it closely matches with the down field sweep (+3 T to 0 T). We have also not shown here the field dependence of $I$, since there has been no considerable change in $I$ with the magnetic field sweep. The $f_r$ at $T$ = 10 K shows a butterfly curve within the field interval of $\mu_0H$ = +1.5 T – -1.5 T and shows a peak on either side of the origin at $\mu_0H$ = ±0.75 T (i.e., $H_P$, the field at which peak occurs) with a large hysteresis, but saturates at the highest field $\mu_0H$ = ±3 T. The field dependent $f_r$ at temperatures above $T$ = 10 K showed the similar behavior as for $T$ = 10 K, except for $T$ = 200 K where the peak is completely vanished with absolutely no hysteresis. The significant point here is that the peak position shifts towards the origin and the hysteresis in $f_r$ becomes narrower at higher temperatures ($T$ > 10 K) as like the hysteresis in $M$-$H$ loops [fig. 2(b)]. This can be clearly concluded from the figures (b) – (e). The magnetic field dependences of $M$ and $f_r$ at $T$ = 10 K are shown in the figures (b) and (c), respectively. A large hysteresis is seen in both the cases. As described above for fig. 6(a), the $f_r$ in the figure (c) showed a peak in the negative side at $H_P$ = -0.75 T while the field is swept from $\mu_0H$ = +3 T → -3 T. This position of the peak ($H_P$) is closely matching with the coercive field ($H_C$), but it is slightly less in value. Interestingly, the temperature dependences of both $H_C$ and $H_P$ show a similar behavior i.e., the rapid increase with lowering temperature below $T = 200$ K and are shown in the figures (d) and (e), respectively.

The resonance frequency of an ICO, $f_r = 1/(2\pi\sqrt{LC})$ where $L$ – the inductance of the empty coil and $C$ – the capacitance in the circuit, changes due to the change in the real part of *rf* magnetic permeability ($\mu'$) of the sample. This is unlike the case of a metallic $La_{0.67}Ba_{0.33}MnO_3$ which showed an extra contribution arising from the incomplete penetration of *ac* magnetic field due to the induced eddy current.[26] Thus, the increase in $\mu'$ upon transition from paramagnetic to ferrimagnetic while cooling leads to a decrease in the $f_r$ around $T_C$. An external applied magnetic field suppresses the spin fluctuations in the sample and leads to a considerable decrease in $\mu'$ which in turn decreases the $f_r$ at $T_C$. On the other hand, the dynamical magnetic and electrical losses in the sample lead to *rf* power absorption in the sample that changes the complex impedance (Z) of the tank circuit, which in turn leads to a change in the current through the ICO circuit. The expression for the impedance of the inductance coil can be modified for a highly insulating materials (in which electromagnetic field completely penetrates the sample) as $Z = (R_0 + \omega L_0 \eta \mu'') + j\omega L_0 \eta \mu'$, where $R_0$ is the resistance, $L_0$ is the inductance of the empty coil, and $\mu'$ and $\mu''$ are the real and imaginary parts of the permeability of the sample, respectively. Therefore, the effective resistance of the coil changes mainly due to the magnetic loss characterized by the $\mu''$ which reflects the electromagnetic power absorption in the sample [$P = \frac{1}{2}H_{ac}^2 \text{Re}(Z)$],[26] and hence the current in the circuit shows an anomaly near the ferrimagnetic transition. The suppression of this peak in the magnetic field of $\mu_0H$ = 60 mT and 0.2 T is due to the suppression of $\mu'$ and $\mu''$ by the magnetic field.

There is a close correlation between the temperature dependence of $H_C$ and $H_P$ [fig. 6 (d) and (e)]. The slight difference in the values of $H_C$ and $H_P$ is possibly because the change in $f_r$ (which is related with the dynamical magnetization i.e. $\delta M/\delta H$) is measured at $f_r \sim$ 1 MHz in our ICO experiment and the low frequency ($f$ <100 kHz) measurement might lead to a closer agreement between the $H_C$ and $H_P$ values. It is worth mentioning that the coercivity of GaFeO$_3$ is high, $H_C$ = 0.9 T at $T$ = 5 K. Since $H_C \propto 2K_1/M_S$ where $M_S$ – saturation magnetization, it implies a rapid increase in the anisotropic constant ($K_1$) with lowering temperature. Indeed, Pinto *et al.*[34] reported unusually large anisotropic magnetization in single crystal of orthorhombic Ga$_{0.92}$Fe$_{1.08}$O$_3$. The coercivity is expected to increase if a Ga is replaced by a rare earth ion R. For example, $H_C$ = 1.5 T, 1.65 T and 1.8 T for R = Dy, Sm and Y, respectively in the RFeO$_3$ series.[35]

## IV. CONCLUSIONS

We have studied electrical, magnetic, magnetodielectric and magnetoabsorption properties of GaFeO$_3$. The dielectric permittivity exhibited a weak anomaly at $T_C$ which is suppressed in $\mu_0H$ = 60 mT suggesting a ME coupling in GaFeO$_3$. The sample showed well defined magnetic hysteresis loops below 230 K with a rapid increase in the coercivity below 200 K from 0.1 T at 200 K to 0.9 T at 5 K. The resistivity of this compound above 200 K obey the Arrhenius law with an activation energy of $E_a$ = 0.67 eV. The *P-E* loops suggested that the leakage current is drastically reduced below $T = 200$ K. The magnetoabsorption study showed an anomaly in both $f_r$ and *I* at $T_C$ and we found that there is a close correlation between the temperature dependence of $H_C$ and $H_P$, the peak found in the $f_r$ versus magnetic field. Very recently, Tokunaga et al.[36] showed that GdFeO$_3$, one of the most orthodox perovskite oxides, possesses a ferroelectric ground state in which ferroelectric polarization is generated by exchange interaction between Gd$^{3+}$ and Fe$^{3+}$ ions. They have also demonstrated the electrical field control of magnetization and the magnetic control of ferroelectric polarization below 2.5 K. In view of these results, it will be interesting to investigate the magnetic and magnetoelectric effects in Ga$_{1-x}$Gd$_x$FeO$_3$.

## ACKNOWLEDGMENTS

R. M. acknowledges the National Research Foundation (Singapore) for supporting this work through the grant NRF-CRP-G-2007.

**References:**

[1] S. W. Cheong and M. Mostovoy, Nat. Mater. **6**, 13 (2007).

[2] R. Ramesh and N. A. Spaldin, Nat. Mater. **6**, 21 (2007).

[3] W. Eerenstein, N. D. Mathur, and J. F. Scott, Nature (London) **442**, 759 (2006).

[4] J. F. Scott, Nat. Mater. **6**, 256 (2007).

[5] M. Gajek, M. Bibes, S. Fusil, K. Bouzehouane, J. Fontcuberta, A. Barthelemy, and A. Fert, Nat. Mater. **6,** 296 (2007).

[6] T. Kimura, T. Goto, H. Shintani, K. Ishizaka, T. Arima and Y. Tokura, Nature **426**, 55 (2003).

[7] D. Lebeugle, D. Colson, A. Forget, M. Viret, A. M. Bataille and A. Gukasov, Phys. Rev. Lett. **100**, 227602 (2008).

[8] Y. Tokunaga, S. Iguchi, T. Arima and Y. Tokura, Phys. Rev. Lett. **101**, 097205 (2008).

[9] K. Taniguchi, N. Abe, S. Ohtani and T. Arima, Phys. Rev. Lett. **102**, 147201 (2009).

[10] N. A. Hill, J. Phys. Chem. B **104**, 6694 (2000).

[11] P. A. Sharma, J. S. Ahn, N. Hur, S. Park, S. B. Kim, S. Lee, J-G Park, S. Guha and S. W. Cheong, Phys. Rev. Lett. **93** (2004) 177202; B. Lorenz, A. P. Litvinchuk, M. M. Gospodinov and C. W. Chu, *ibid*, **92** (2004) 087204; T. Goto, T. Kimura, G. Lawes, A. P. Ramirez and Y. Tokura, *ibid*, **92** (2004) 257201.

[12] J. P. Remeika, J. Appl. Phys. **31**, 263S (1960).

[13] G. T. Rado, Phys. Rev. Lett. **13**, 335 (1964).

[14] D. N. Astrov, J. Exptl. Theoret. Phys. (U.S.S.R.) **38**, 984 (1960) [translation: Soviet Phys. JETP **11**, 708 (1960)]; V. J. Folen, G. T. Rado and E. W. Stalder, Phys. Rev. Lett. **6**, 607 (1961).

[15] M. P. Petrov, S. A. Kizaev and G. A. Smolenskyi, Solid State Commun. **8**, 195 (1967).

[16] V. J. Folen and G. T. Rado, Solid State Commun. **7**, 433 (1969).

[17] Y. Kaneko, T. Arima. J. P. He, R. Kumai and Y. Tokura, J. Magn. Magn. Mater. **272-276**, 555 (2004).

[18] E. A. Wood, Acta Cryst. **13**, 682 (1960).

[19] R. B. Frankel, N. A. Blum, S. Foner, A. J. Freeman and M. Schieber, Phys. Rev. Lett. **15**, 958 (1965).

[20] S. C. Abrahams and J. M. Reddy, Phys. Rev. Lett. **13**, 688 (1964).

[21] Y. Ogawa, Y. Kaneko, J. P. He, X. Z. Yu, T. Arima and Y. Tokura, Phys. Rev. Lett. **92**, 047401 (2004).

[22] J. H. Jung, M. Matsubara, T. Arima, J. P. He, Y. Kaneko and Y. Tokura, Phys. Rev. Lett. **93**, 037403 (2004); N. Kida, Y. Kaneko, J. P. He, M. Matsubara, H. Sato, T. Arima, H. Akoh and Y. Tokura, Phys. Rev. Lett. **96**, 167202 (2006).

[23] T. Arima, D. Higashiyama, Y. Kaneko, J. P. He, T. Goto, S. Miyasaka, T. Kimura, K. Oikawa, T. Kamiyama, R. Kumai and Y. Tokura, Phys. Rev. B **70**, 064426 (2004).

[24] M. Matsubara, Y. Kaneko, J.-P. He, H. Okamoto and Y. Tokura, Phys. Rev. B **79**, 140411 (R) (2009).

[25] A. M. Kalashnikova, R. V. Pisarev, L. N. Bezmaternykh, V. L. Temerov, A. Kirilyuk and Th. Rasing, JETP Lett. **81**, 452 (2005).

[26] V. B. Naik and R. Mahendiran, Appl. Phys. Lett. **94**, 142505 (2009).

[27] S. C. Abrahams, J. M. Reddy and J. L. Bernstein, J. Chem. Phys. **42**, 3957 (1965).

[28] T. Kimura, S. Kawamoto, I. Yamada, M. Azuma, M. Takano and Y. Tokura, Phys. Rev. B. **67**, 180401(R) (2003).

[29] Z. H. Sun, B. L. Cheng, S. Dai, L. Z. Cao, Y. L. Zhou, K. J. Jin, Z. H. Chen and G. Z. Yang, J. Phys. D: Appl. Phys., **39**, 2481 (2006).

[30] Y. Aikawa, T. Katsufuji, T. Arima and K. Kato, Phys. Rev. B. **71**, 184418 (2005).

[31] T. Kimura, Y. Sekio, H. Nakamura, T. Siegrist and A. P. Ramirez, Nat. Mater. **7,** 291 (2008).

[32] B. Lorenz, Y. Q. Wang, Y. Y. Sun and C. W. Chu, Phys. Rev. B. **70**, 212412 (2004).

[33] M. Dawber, K. M. Rabe, J. F. Scott, Rev. Mod. Phys. **77**, 1083 (2005).

[34] A. Pinto, J. Appl. Phys. **37**, 4372 (1966).

[35] D. S. Schmool, N. Keller, M. Guyot, R. Krishnan and M. Tessier, J. Magn. Magn. Mater. **195**, 291 (1999).

[36] Y. Tokunaga, N. Furukawa, H. Sakai, Y. Taguchi, T. Arima and Y. Tokura, Nat. Mater. **8**, 558 (2009).

**Figures :**

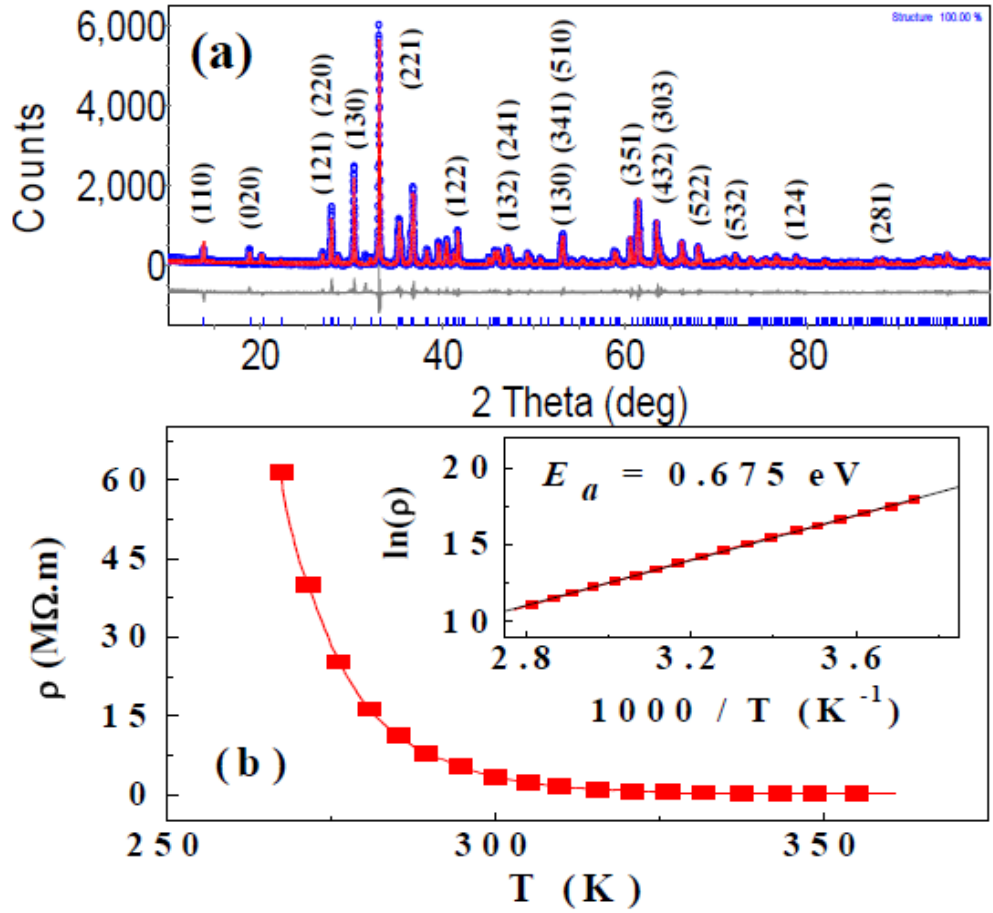

Fig. 1 (color online) (a) Observed (blue color) and Reitveld refinement (red color) of the XRD pattern for the GaFeO$_3$ compound with space group *Pc2$_1$n* at room temperature, (b) temperature dependence of the resistivity ($\rho$) in a narrow temperature interval (265 K – 360 K) and the inset shows the linear behavior of Arrhenius plot for the resistivity.

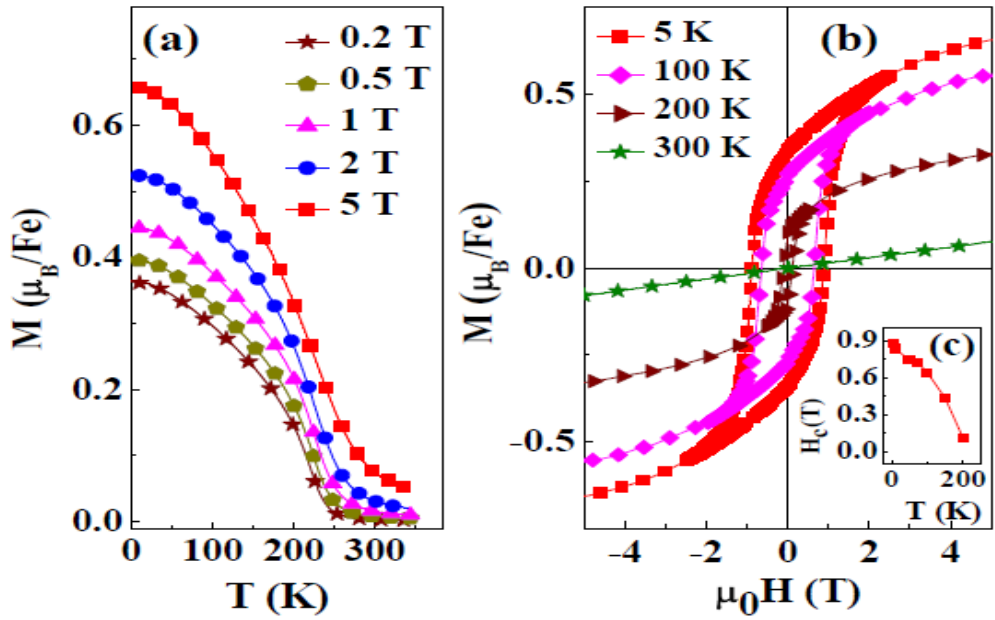

**Fig. 2 (color online) (a) Temperature dependences of magnetization (*M*) at different magnetic fields (0.2 T – 5 T), (b) field dependences of magnetization (*M*) at different temperatures (5 K – 300 K), (c) temperature dependence of coercive field (*H*$_C$).**

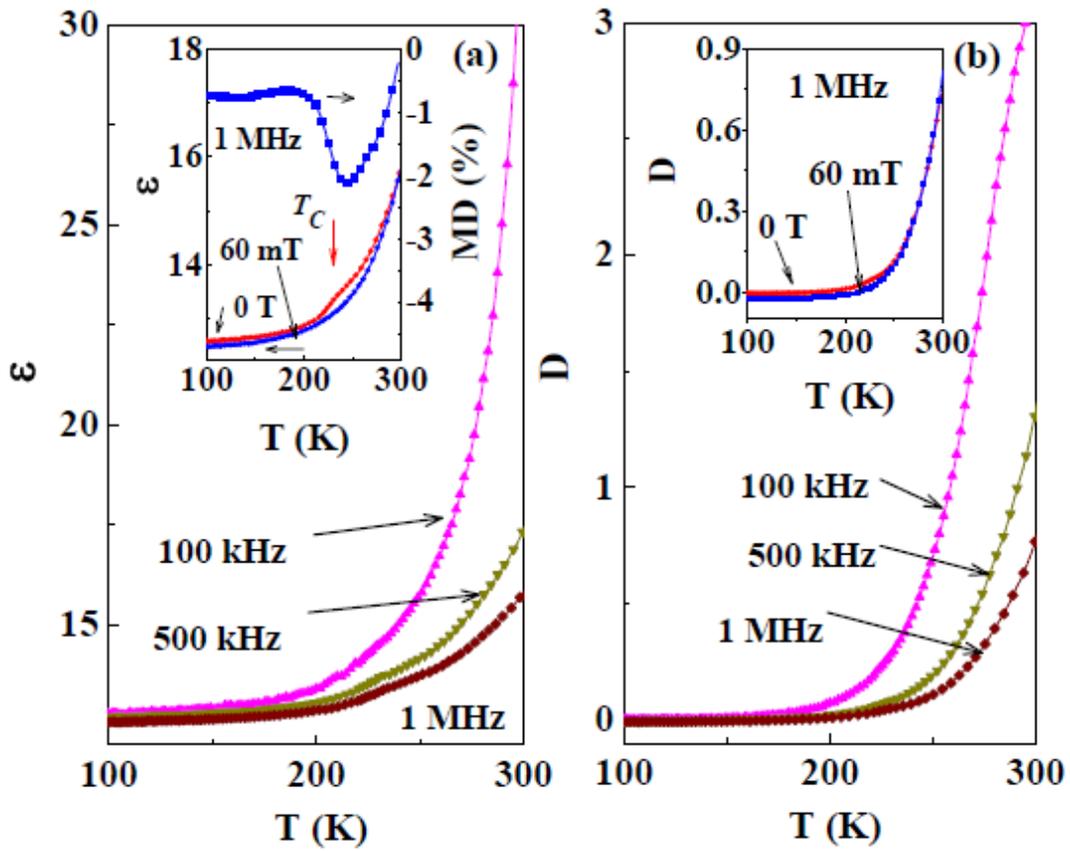

**Fig. 3 (color online)** Temperature dependences of (a) the dielectric constant ($\varepsilon$) and (b) the dissipation factor ($D = \tan\delta$) for $f = 100$ kHz, 500 kHz and 1 MHz. The insets show the temperature dependences of $\varepsilon$ and $D$ (left scale) in $\mu_0 H = 0$ and 60 mT for $f = 1$ MHz. The inset in the figure (a) shows the magnetodielectric (MD) coefficient (right scale) as a function of temperature for $f = 1$ MHz.

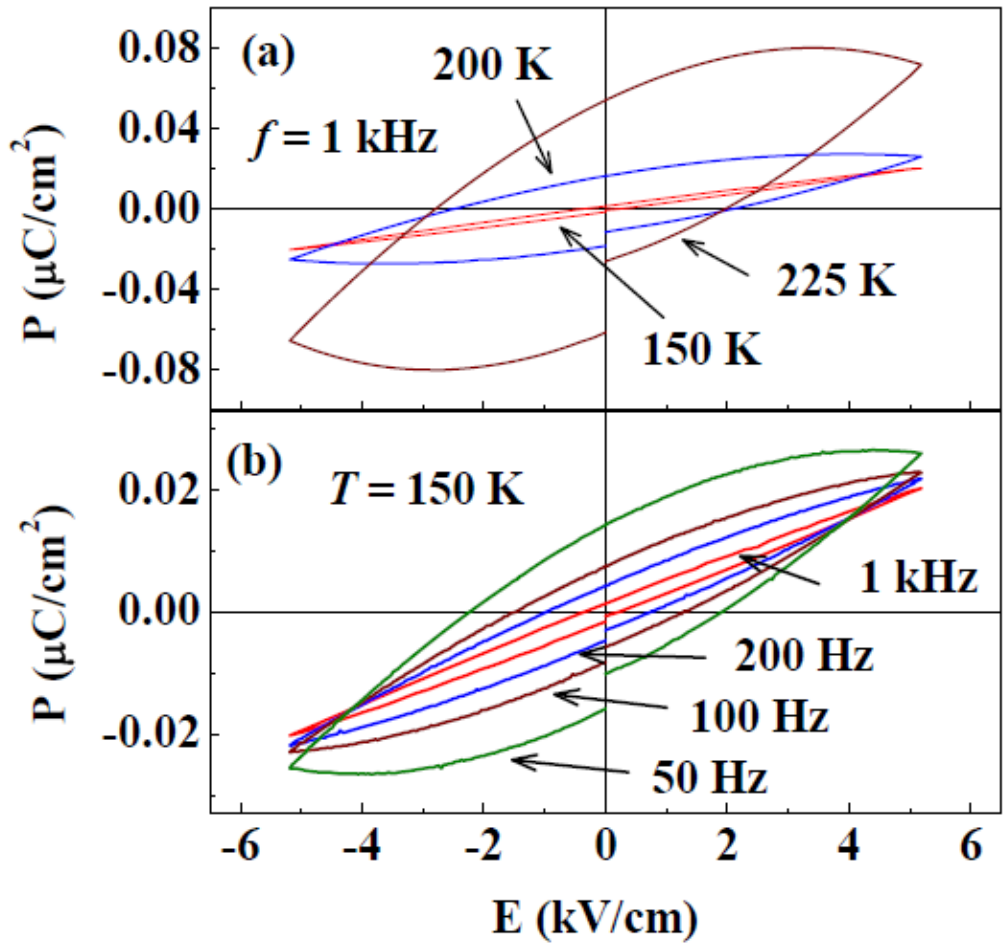

**Fig. 4 (color online) (a) *P-E* loops at selected temperatures (frequency of the hysteresis cycle *f* = 1 kHz) (b) *P-E* loops at *T* = 150 K with different frequencies of the hysteresis cycle (*f* = 1 kHz – 50 Hz).**

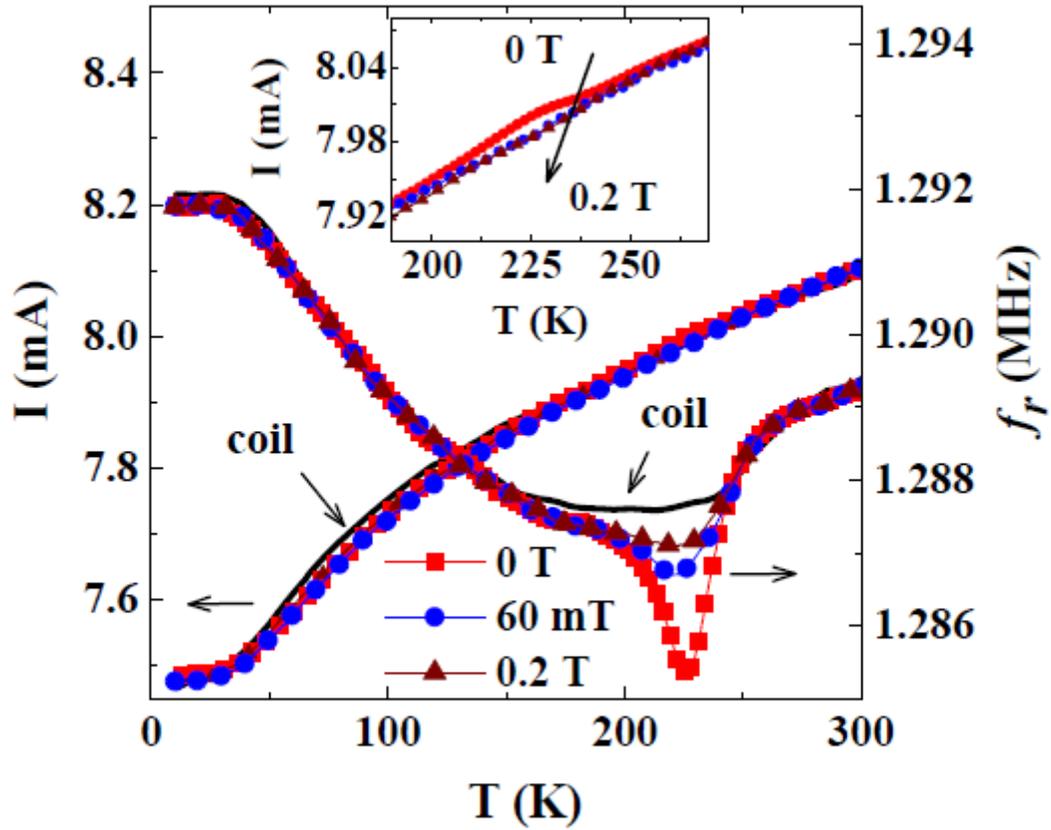

**Fig. 5 (color online) Temperature dependences of the resonance frequency ($f_r$) (right scale) and current (*I*) (left scale) through the circuit for different strengths of *dc* magnetic fields $\mu_0 H$ = 0 T, 60 mT and 0.2 T. The data for the empty coil are also shown. The inset shows the temperature dependence of *I* in a narrow temperature interval (190 K – 270 K).**

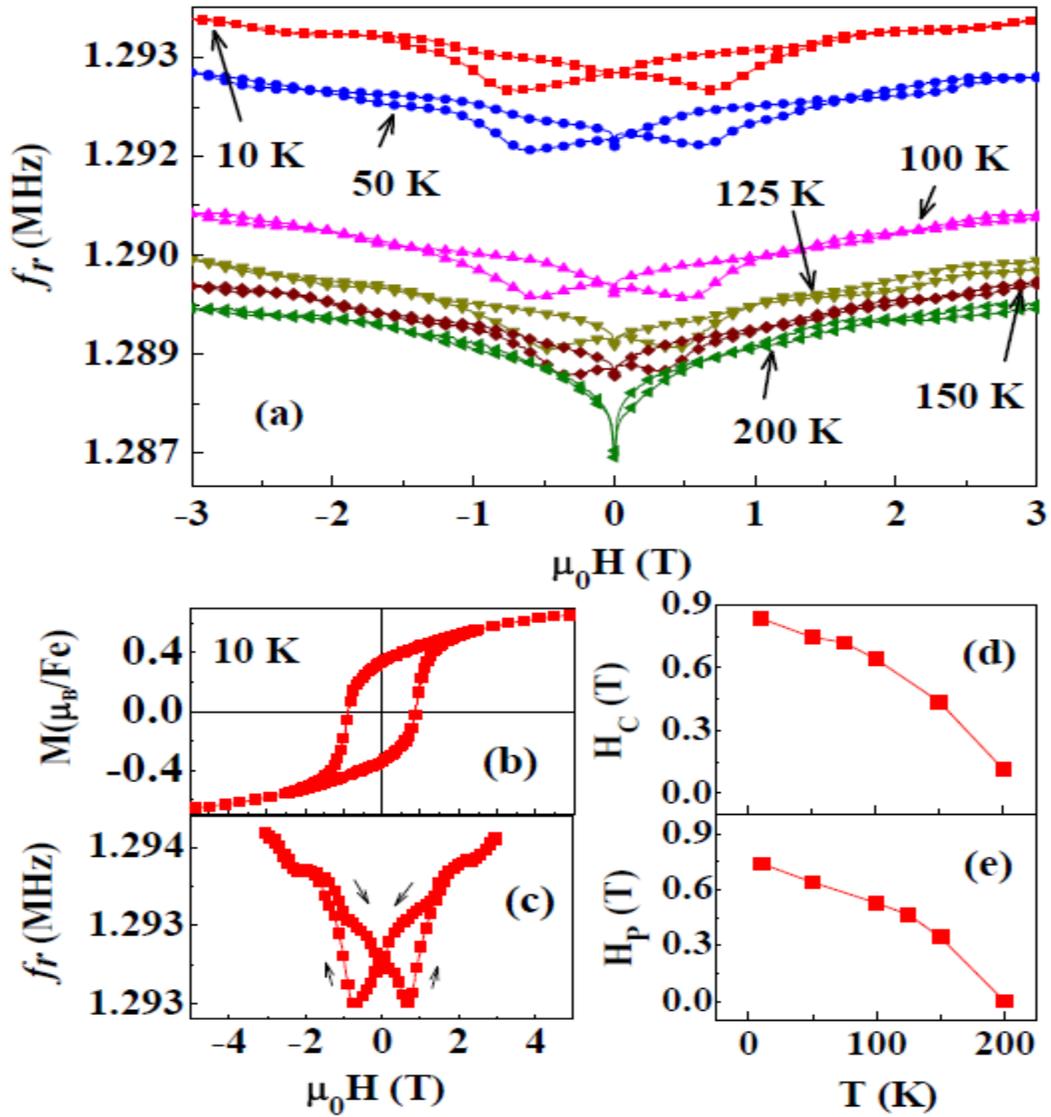

Fig. 6 (color online) The magnetic field dependences of the (a) resonance frequency ($f_r$) at selected temperatures $T$ = 10 K – 200 K, (b) magnetization ($M$) at $T$ = 10 K, (c) resonance frequency ($f_r$) at $T$ = 10 K. Temperature dependences of the (d) coercive field ($H_C$) and (e) position of the peak ($H_P$) which is observed in the $f_r$ versus field.